\documentclass[prd,twocolumn,nofootinbib,amsfonts]{revtex4-2}
\usepackage[utf8]{inputenc}
\usepackage[T1]{fontenc}
\usepackage[colorlinks, linkcolor = red]{hyperref}
\usepackage{graphicx,bm,amsmath,scalerel}
\usepackage{bbold}
\usepackage{wasysym}
\usepackage{verbatim}
\usepackage{amssymb}
\usepackage{epstopdf}
\usepackage{xmpmulti}
\usepackage{centernot}
\usepackage{multirow}
\usepackage{tikz}
\usepackage{graphicx}
\usepackage{float}
\usepackage{mathtools}
\usepackage{comment}

\newcommand{\dd}{{\rm d}}

\makeatletter

\newcommand{\be}{\begin{equation}}
	\newcommand{\ee}{\end{equation}}
\newcommand{\beq}{\begin{eqnarray}}
	\newcommand{\eeq}{\end{eqnarray}}
\newcommand{\ba}{\begin{align}}
	\newcommand{\ea}{\end{align}}

\usepackage{xcolor}
\begin{document}

    \title{Exploring Born-Infeld f(T) teleparallel gravity \\through accretion disk dynamics}

	\author{Ruijing Tang $^{1,2,3,4}$ 
      \email{r3tang@uwaterloo.ca}, Shokoufe Faraji $^{1,2,3}$ 
          \email{s3faraji@uwaterloo.ca}, Niayesh Afshordi$^{1,2,3}$ \email{nafshordi@pitp.ca}  }
  \affiliation{$^{1}$Department of Physics and Astronomy, University of Waterloo, Waterloo, Canada}
  \affiliation{$^{2}$Waterloo Centre for Astrophysics, University of Waterloo, Waterloo, Canada}
 \affiliation{$^{3}$Perimeter Institute for Theoretical Physics, Waterloo, Canada}
 \affiliation{$^{4}$Observatoire de Paris, Universite PSL, Paris, France}
% % \affiliation{Department of Physical Science and Engineering, Beijing Jiaotong University, Beijing 100044, China}

\begin{abstract}

% The main goal of this paper is to investigate the properties of accretion into non-spinning black holes in Teleparallel Born-Infeld gravity. This study aims to assess the applicability of this modified gravity theory in observational settings by analyzing its impact on accretion disk properties. The results could provide insights into the viability of Teleparallel Born-Infeld gravity as an alternative to general relativity in describing strong gravitational fields and accretion phenomena.
Teleparallel Born-Infeld gravity (TBI) is a modified theory of gravity that aims to maintain second-order field equations, leading to alternative scenarios for strong gravity and cosmological settings. In this study, we examine the impact of TBI gravity on the physical characteristics of thin (Novikov-Thorne) accretion disks, focusing on quantities such as flux, pressure, temperature, etc. We also examine the spectral luminosity, comparing it to disks around the Schwarzschild black holes. By comparing the theoretical predictions to observational data in the low frequency regime, we demonstrate the model’s ability to match real astrophysical systems and distinguish subtle differences between TBI gravity and general relativity, with improved sensitivity. Furthermore, the results suggest that observations of X-ray spectra from the inner disk regions can provide valuable insights into the properties of TBI gravity, potentially offering constraints on this modified gravity theory through future astrophysical observations.
\end{abstract}

\maketitle

%\tableofcontents

%%%%%%%%%%%%%%%%%%%%%%%%%%%%%%%%%%%%%%%%%%%%%%%%%%%%%%%%%%%%%%%%%%%%%%%%%%%%%%%%%%%%%%%%%%%%%%%%%%%%%%%%%%%%%%%%%%%%%%%%%%%%%%%%%%%%%%%%%%
\section{Introduction}
The theory of black hole accretion disks is a significant area of research within fundamental physics, as accretion disks play a crucial role in high-energy astronomical phenomena. Accretion disks are believed to be responsible for the immense energy outputs observed in various astrophysical systems and galaxies because they can extend deep into the strong gravitational fields of black holes and compact objects e.g., \cite{1969Natur.223..690L,1995cvs..book.....W,2013A&ARv..21...69R}. This unique characteristic allows them to serve as natural laboratories for testing the predictions of general relativity and its numerous modifications and extensions. In fact, given the limitations and challenges faced by general relativity in extreme conditions (namely singularities and lack of a consistent quantum theory), it becomes necessary to explore modified gravity theories. These alternative theories aim to address the gaps and extend our understanding of gravitational interactions beyond the framework of general relativity. By studying the properties and behaviors of accretion disks under these modified theories, we can test their validity to explain astrophysical phenomena. This area of research is crucial for developing a more comprehensive theory of gravity that can explain observations across all scales and conditions in the universe.

%%%%%%%%%%%%%%%%%%%%%%%%%

%%%%%%%%%%%%%%%% TBI%%%%%%%%%%%%%%%%%

In recent years, there has been a growing interest in modified gravity theories to address and potentially resolve certain puzzling aspects of standard gravitational theory and cosmology, such as dark matter,  cosmic inflation, cosmic singularities, the particle horizon problem, and the accelerated expansion of the Universe. By introducing variations and extensions to general relativity, we hope to gain deeper insights into the fundamental nature of gravity as well as the small-scale and large-scale structure of the Universe. Many of these modified theories of gravity involve simple deformations in the equations or principles of general relativity theory such as additional spatial dimensions, modifying the action, adding new fields that interact with gravity to provide new dynamics or incorporating effects from quantum mechanics. %For example, the Lovelock Lagrangian is constructed as a polynomial in the Riemann curvature tensor, which leads to second-order field equations for the metric tensor \cite{10.1063/1.1665613}. However, it deviates from the Einstein-Hilbert Lagrangian only in spacetimes with more than four dimensions. 
For instance, Lovelock gravity generalizes general relativity to higher dimensions by including higher-order curvature terms in action. By its construction, the equations of motion remain second-order in the derivatives of the metric \cite{Lovelock:1971yv}. Nevertheless, Lovelock equations of motion only differ from Einstein equations, for dimension larger than four. In contrast, $f(R)$ theories that primarily aim to explain cosmic acceleration without relying on quintessence models \cite{1982GReGr..14..453K}, lead to fourth-order differential equations in the metric tensor,  since the action contains a general function of the Ricci scalar instead of being directly proportional to it.

One approach to obtaining second-order equations in four dimensions is to utilize the Teleparallel Equivalent of General Relativity (TEGR) rather than the Einstein-Hilbert Lagrangian. While general relativity employs the Levi-Civita connection, which features curvature without torsion, teleparallelism uses the Weitzenböck connection, which has zero curvature but nonzero torsion. This torsion is what accounts for the gravitational interaction. In this framework, teleparallelism  can be viewed as a subset of Einstein-Cartan theories \cite{RevModPhys.48.393}, which describe gravity through a connection that encompasses both torsion and curvature. In teleparallel gravity, gravity is interpreted as a force, similar to the other fundamental forces, and it acts through torsion. This is in contrast to GR, where gravity is seen as the effect of spacetime curvature. TEGR is a specific form of teleparallel gravity that is equivalent to general relativity at the level of field equations. It uses a tetrad (or vierbein) formalism. Teleparallel gravity provides a different but equivalent framework to general relativity for describing the gravitational interaction, for example, on energy-momentum localization and can be extended to form the basis of modified gravity theories, such as $f(T)$ gravity. In particular, TEGR admits an interpretation as a gauge theory of the translation group, where the tetrad (equivalently, a translational gauge potential) plays the role of the gauge field and torsion is the corresponding field strength (see e.g. \cite{AldrovandiPereiraVu2004,AldrovandiPereira2013}). On the other hand, Born-Infeld gravity theory is inspired by Born-Infeld electrodynamics, which was initially formulated to address the problem of infinite self-energy of point charges in classical electrodynamics \cite{1934RSPSA.144..425B}. This approach was extended to gravity to regularize singularities. Interestingly, Teleparallel Born-Infeld gravity (TBI) combines ideas from teleparallel gravity and Born-Infeld type modifications Teleparallel Born-Infeld deformations of TEGR were originally proposed in ~\cite{FerraroFiorini2007,FerraroFiorini2008} and subsequently developed in a variety of cosmological and strong field applications (see e.g, \cite{FioriniFerraro2009,PhysRevD.86.083515,Jana2014,JusufiEtAl2022}). As a result, in this theory, the action is based on the Weitzenb\"ock connection (in contrast to the Levi-Civita connection used in general relativity) and includes a Born-Infeld-like structure. The torsion scalar $T$ replaces the Ricci scalar $R$ in the standard Born-Infeld gravity action. The field equations are derived from this modified action, resulting in a different set of equations compared to standard teleparallel gravity and Born-Infeld gravity. The torsion tensor, rather than the curvature tensor, plays a central role. In addition, in TBI the equations of motion remain second-order, thereby avoiding higher-derivative (Ostrogradsky-type) instabilities.\footnote{This property alone does not guarantee the absence of additional propagating modes, background-dependent branching of degrees of freedom, or strong-coupling behavior in generic $f(T)$-type modifications; see e.g. \cite{GolovnevGuzman2020,DelhomJimenezCanoMaldonadoTorralba2023} and references therein.}

 % In addition, in TBI the equations of motion remain second-order \footnote{{\color{blue}Second order field equations avoid the Ostrogradsky-type instabilities typically associated with higher-derivative actions. This property alone does not guarantee the absence of additional propagating modes or strong-coupling behavior in generic $f(T)$-type modifications; see e.g. \cite{GolovnevGuzman2020,DelhomJimenezCanoMaldonadoTorralba2023} and references therein.}}. 

For these reasons, TBI gravity offers a proper instructive testbed. This model preserves the second order dynamics characteristic of the $f(T)$ class, introduces only one additional scale via its Born-Infeld parameter, which can moderate curvature or torsion divergences in various branches and, crucially for the present study, possesses an exact spherically symmetric black hole solution. The availability of that analytic metric enables offers a rich phenomenology that predicts potential observable effects that can be tested via observational data within a torsion based framework.

% Similarly to how Born-Infeld electrodynamics regularizes the infinite self-energy of point charges, TBI gravity aims to regularize singularities \cite{Ferraro:2006jd,Bohmer:2019vff,2020arXiv200511843B} in the gravitational field, which can lead to more physically realistic models of compact objects. Furthermore, TBI gravity may provide novel alternative strong gravity and cosmological scenarios. Its theoretical appeal and practical advantages make it a compelling area of research for physicists exploring the fundamental nature of gravity, since this theory offers a rich phenomenology that predicts a variety of potential observable effects that can be tested via observational data.

Continuing the investigation of equilibrium configurations of thick disk models without active accretion in this spacetime \cite{2022PhRvD.106h4046B}, this work shifts the focus to thin accretion disks, where the presence of accretion plays a key role in their dynamics. Studying thin disks in this gravitational background provides new insights into accretion processes and enhances our understanding of disk behavior in modified gravity frameworks. The standard thin accretion disk model, first laid out by Bardeen, Press and Teukolsky \cite{1972ApJ...178..347B}, Shakura and Sunyaev \cite{1973A&A....24..337S}, Novikov and Thorne \cite{1973blho.conf..343N}, and Lynden-Bell and Pringle \cite{1974MNRAS.168..603L}, provides a fundamental framework for understanding the behavior of matter as it accretes onto compact objects. In this model, the disk is geometrically thin and optically thick, meaning it radiates efficiently while maintaining a relatively small vertical height compared to its radial extent.

%%%%%%%%%%%%%%% content %%%%%%%%%

The structure of this paper is as follows. In Section \ref{sec:TBI} we briefly introduce TBI gravity and its background geometry. The thin disk model is briefly described in Section \ref{sec:disk}. The properties of the disk in this background are computed in Section \ref{sec:Result}. We summarize our results and conclude in Section \ref{sum}. In this paper, we adopt the metric signature $(-,+,+,+)$ and use geometric units where $c=G=1$, unless otherwise stated. Additionally, an over-dot denotes differentiation with respect to the affine parameter, while a prime denotes differentiation with respect to the radial coordinate.

%%%%%%%%%%%%%% Spacetime %%%%%%%%%%%%

\section{Teleparallel Born-Infeld gravity}\label{sec:TBI}

The action for Teleparallel Born-Infeld (TBI) gravity is given by
\begin{align}
S_{\rm TBI} = \frac{1}{2\kappa} \int   e f(T) \, d^4x,
\end{align}
where $f(T)$ is given by

\begin{align}\label{eq:TBIact}
    f(T)=  \tilde{\Lambda} \left( \sqrt{1 + \frac{2T}{\tilde{\Lambda}}} - 1 \right),
\end{align}
here $e = \det(e^A_\mu)$ is the determinant of the tetrad fields which are the fundamental variables in teleparallel gravity, $T$ is the torsion scalar, $\kappa = 8 \pi G$ is the gravitational constant, and $\tilde{\Lambda}$ is a parameter introducing the Born-Infeld non-linearity. The specific form of this action introduces a non-linear modification that regularizes the action. In the limit $\tilde{\Lambda} \to \infty $ the function
$f(T)$ reduces to $T$, recovering the standard TEGR, which is equivalent to general relativity. 

\begin{figure}
    \centering
    \includegraphics[width=\hsize]{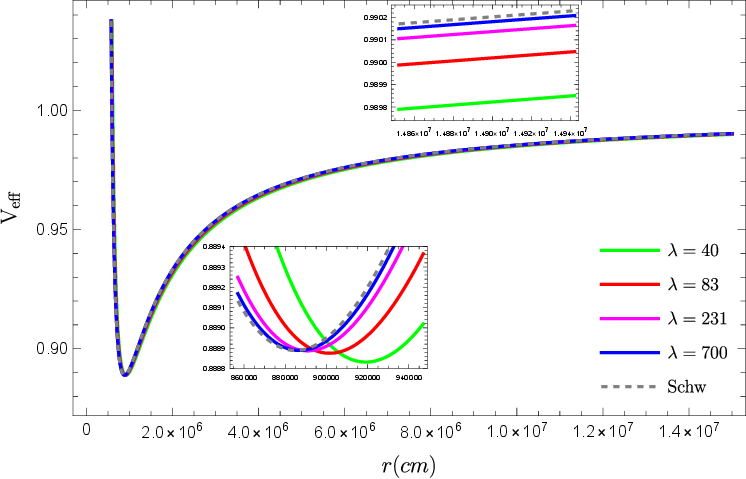}
    \caption{Effective potential for massive particle orbiting a black hole, in general relativity ($\lambda \to \infty$) and TBI gravity (finite $\lambda$).}
    \label{fig:Veff}
\end{figure}

TBI gravity defined by \eqref{eq:TBIact}, has the advantage of admitting an analytical solution in spherical symmetry \cite{Bahamonde:2021srr}. Although \cite{Bahamonde:2021srr} is formulated in the broader $f(T,B)$ framework, the specific background employed here is obtained there in the $B$-independent sector, i.e.\ as an exact vacuum solution of the Born-Infeld $f(T)$ model \eqref{eq:TBIact} (complex-tetrad branch;  \cite[see Eqs.~(100-101) and the surrounding discussion]{Bahamonde:2021srr}). For convenience, we use the equivalent $A(r)$-$B(r)$ parametrization (see also \cite{2022PhRvD.106h4046B}). The metric reads as follows

\begin{align}\label{eq:fTBImetric}
    \dd s^2 = - A(r) \dd t^2 +  \frac{B(r)}{A(r)}\dd r^2 + r^2\dd \Omega^2\,,
\end{align}
where 
\begin{align}
    A(r) & :=1-\frac{2 M}{r}-\frac{2 M}{r \lambda} \tan ^{-1}\left(\frac{\lambda  r}{2 M}\right) \,, \label{defA}\\
    B(r) & :=\frac{r^4 \lambda^4}{16 M^4 \left(1+\frac{\lambda ^2 r^2}{4 M^2}\right)^2}\,,
\end{align}
and $\lambda=M\sqrt{\tilde{\Lambda}}$ \footnote{In modified teleparallel gravity the fundamental variables are the tetrad $e^{A}{}_{\mu}$ (and, in the covariant
formulation, an inertial spin connection), so a spacetime solution should be understood as a consistent tetrad-(spin-connection)
pair rather than as a metric alone. The line element \eqref{eq:fTBImetric} corresponds to the spherically symmetric Weitzenb\"ock($\omega^{A}{}_{B\mu}=0$) complex-branch tetrad given in \cite[Eq. 40]{Bahamonde:2021srr}, which satisfies the antisymmetric (spin-connection) field equations; for the Born-Infeld model this yields the exact solution.}. Using the Parameterized Post-Newtonian (PPN) formalism pioneered by Kenneth Nordtvedt \cite{PhysRev.169.1017}, we can place lower limits on the parameter $\tilde{\Lambda}$. The PPN framework allows us to express the weak-field predictions of the theory in terms of standardized PPN parameters, such as $\gamma$ and $\beta$ \footnote{The parameter $\gamma$ measures the amount of space curvature produced by unit rest mass and the parameter $\beta$ measures the nonlinearity in the superposition law for gravity,  quantifying the degree of self-interaction of the gravitational field.}. To analyze the weak field limit, we expand the  metric for large $\frac{r}{M}$ \cite{2022PhRvD.106h4046B}

\begin{align}
  - g_{tt} & = 1 - \frac{2M}{r} \left( 1 + \frac{\pi}{2\lambda} \right) + \frac{4M^2}{r^2 \lambda^2} + \mathcal{O}(r^{-4})\,,\\
    g_{rr} & = 1 + \frac{2M}{r} \left( 1 + \frac{\pi}{2\lambda} \right) + \mathcal{O}(r^{-2}).
\end{align}
and compare this expression to the standard PPN expansion,
\begin{align}
   - g_{tt} & = 1 - \frac{2\hat{M}}{r} + (\beta-\gamma) \frac{2\hat{M}^2}{r^2}\\
     g_{rr} & = 1 + \gamma \frac{2\hat{M}}{r},
\end{align}
where $\hat{M} = Gm/c^2$ and $m$ is the Newtonian mass, we find consistently that $\hat{M}=M_{\textrm{Komar}}= \frac{1}{2}\lim_{r\to\infty}\left( r^2 \frac{g_{tt}'}{g_{tt}} \right) = \left(\frac{\pi }{2\lambda}+1\right) M$. Adopting the Cassini bound on $\gamma$, these analyses yield a bound on $\beta$ given by agrees with its general relativity value
\begin{align}
    (\beta-1) &= \frac{8}{\pi^2} \Big(1-\frac{M}{\hat{M}}\Big)^2 = \frac{8}{(2\lambda+\pi)^2},
\end{align}
using the observational bound from the perihelion shift of Mercury: $(\beta-1)=(-4.1 \pm 7.8)\times 10^{-5}$   \cite{Will:2014kxa}, from the perihelion advance of Mars: $(\beta-1)=(0.4 \pm 2.4)\times 10^{-4}$ \cite{2011Icar..211..401K}, from INPOP08: $(\beta-1)=(0.75 \pm 1.25)\times 10^{-4}$, and from INPOP10a: $(\beta-1)=(-0.62 \pm 0.81)\times 10^{-4}$ \cite{2011CeMDA.111..363F}, we obtain different lower bounds on $\tilde{\Lambda}$ respectively as 

\begin{align}\label{eq:lambdappn}
    \lambda_{\rm PPN}= M_\odot \sqrt{\tilde{\Lambda}} > 231, 83, 98, 323.
\end{align}
To study the particle motion in this spacetime especially near the innermost stable circular orbit (ISCO), considered the inner edge of a thin accretion disk, we need to analyze the behaviour of effective potential. Following the standard procedure for deriving the radial component of the geodesic motion, we obtain 
\begin{align}
    \dot r^2 = \frac{1}{B(r)} \left(E^2 - V_{\rm eff}(r)\right)\,,
\end{align}
where $E$ is conserved energy per unit mass of the particle and $V_{\rm eff}$ is effective potential

\begin{align} \label{effparticle}
    V_{\rm eff} & = A(r) \left(1+\frac{L^2}{r^2}\right)\,.
\end{align}
and $L$ is the angular momentum per unit mass of the particle. Figure \ref{fig:Veff} shows the behaviour of $V_{\rm eff}$ for some chosen parameters of $\lambda$. As $\lambda$ decreases, the deviation from the Schwarzschild solution becomes more pronounced. Note that Some figures include $\lambda$ values below the weak-field bound in equation \eqref{eq:lambdappn} purely to enhance visual separability. Since the leading corrections scale as $\mathcal{O}(\lambda^{-1})$, curves in the allowed regime would overlap at the plotted scale.

Notably, the minimum of $V_{\rm eff}$ as given by \eqref{effparticle}, determines the ISCO for massive particles. Consequently, this minimum, which also marks the inner edge of the accretion disk, occurs at a larger radius for smaller values of $\lambda$. The criteria for a circular orbit are met when both the radial velocity and radial acceleration are zero, meaning $\dot{r}=0$, and $\ddot{r}=0$. Thus, to the first order in $\lambda^{-1}$ we obtain
\begin{align}\label{rms}
    r_{\rm ISCO} & = 6 + \frac{3\pi}{\lambda}\,.
\end{align}
In the following section, we will employ this equation to determine the edge of the thin accretion disk as we investigate the disk properties within this background. This worth emphasizing that in this work we treat this geometry as a background and focus on thin disk observables.
A dedicated perturbation analysis of this black hole solution (e.g. dynamical stability and the propagation of additional modes)
is beyond our scope; the disk signatures derived below should therefore be interpreted as conditional predictions for a long-lived
astrophysical realization of the spacetime.

%%%%%%%%%%%%%%%%%%%%%%%%%%%%%%%%%%%%%%%%%%%%%%%%%%%%%%%%%%%%%%%%%%%%%%%%%%%%%%%%%%%%%%%%%%%%%

\section{Thin accretion disk model} \label{sec:disk}

In the study of accretion disks, the standard thin accretion disk model plays a pivotal role. This model assumes a steady axisymmetric fluid configuration where the physical quantities depend on both the vertical distance from the equatorial plane and the radial distance from the central object. The thin disk model assumes that the disk is razor-thin and confined to the equatorial plane, with a small ratio of disk half-thickness $H(r)$ to radius $r$, $H/r \ll 1$. This ensures that heat generation and radiation losses are balanced and leads to negligible advection. Consequently, the disk's luminosity is typically below 30\% of the Eddington luminosity, and the mass accretion rate $\dot{M}$ stays below the Eddington rate $\dot{M}_{\rm Edd}$. Beyond this threshold, the gas becomes too optically thick to radiate all the dissipated energy locally \cite{2006ApJ...652..518M,2008ApJ...676..549S}, making it difficult to justify the use of the standard thin-disk model at higher luminosities. Given the geometrically thin assumption, the disk's two-dimensional structure can be simplified into radial quasi-Keplerian flow and vertical hydrostatic balance.

A key feature of the thin disk model is its ability to locally radiate a large fraction of its rest mass energy as near-thermal black-body radiation, driven by viscous mechanisms. This causes the thin model to be classified as a cold accretion disk (compared to the virial temperature). However, this model does not explain the very high temperatures ($T>10^{10} K$) observed at the Galactic Center source Sgr A * \cite{2007MNRAS.381.1267K}, which nowadays is modeled by a magnetically arrested disk \cite{2003PASJ...55L..69N}. Additionally, a thin disk model assumes that the specific internal energy density is negligible, and the disk lies in the equatorial plane, implying the $u^{\theta}$ component of the fluid's four-velocity vanishes. Furthermore, quasi-Keplerian circular orbits with a small radial drift velocity $u^{r}$ are also assumed. The inner edge of a thin accretion disk with sub-Eddington luminosities is typically at the ISCO, where most of the luminosity originates. In this model, shear stress, attributed to a form of viscosity, is responsible for transporting angular momentum and energy outward while accreting matter inward. This model introduces viscosity through a so-called $\alpha$-prescription without specifying the viscosity concept. However, the viscosity in the accretion process cannot be the same as molecular viscosity and may have a magnetic nature \cite{1991ApJ...376..214B,1992ApJ...400..610B}. The model suggests that shear stress functions as a type of viscosity, which is crucial for transporting angular momentum and energy outward, driving matter inward for accretion, and locally heating the gas. Therefore, in this model, the shear stress is proportional to the total pressure, with the dimensionless parameter $\alpha$ acting as the proportionality coefficient \cite{1973A&A....24..337S}.

\begin{figure*}
    \centering
    \includegraphics[width=0.49\hsize]{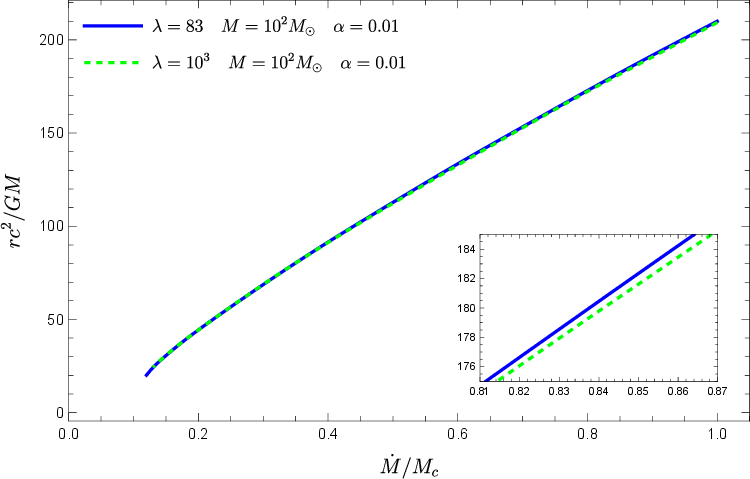}
    \includegraphics[width=0.49\hsize]{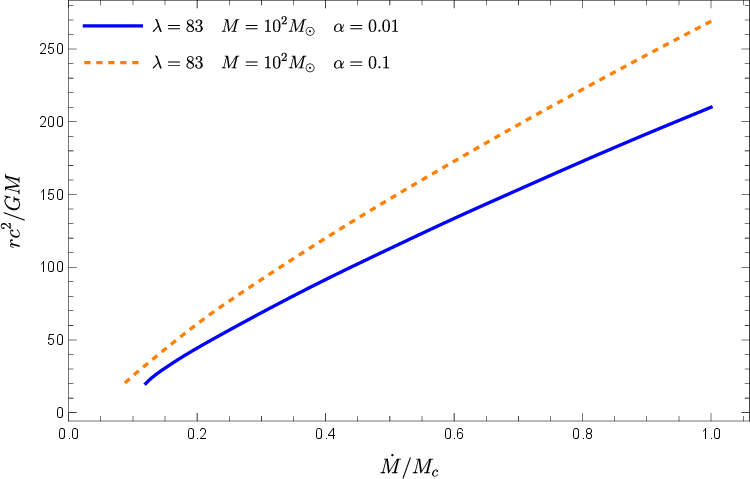}
    \includegraphics[width=0.49\hsize]{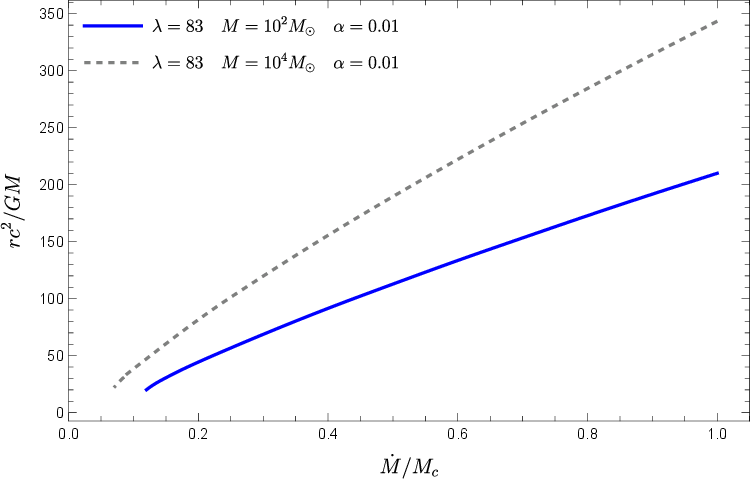}
    \includegraphics[width=0.49\hsize]{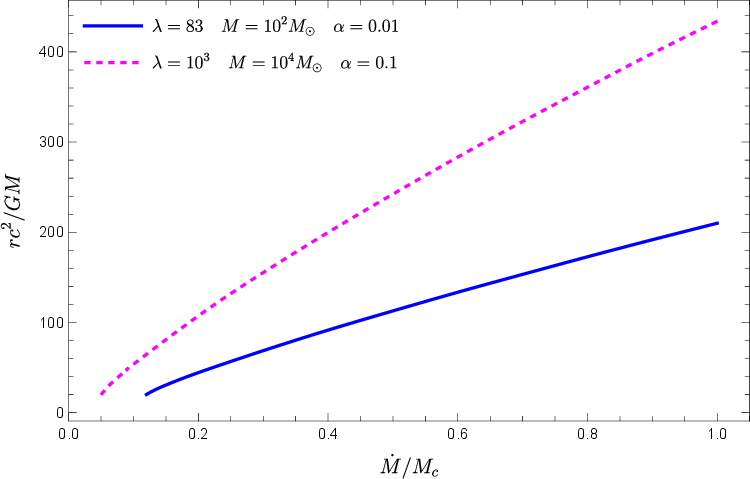}
    \caption{Radius at which $P_{\rm gas}/P_{\rm rad}=1$. \label{fig:startingp-comparison3}}
\end{figure*}

\subsection{Main Equations}
The radial structure of the thin disk model is governed by three fundamental equations. The first equation is the conservation of particle number, expressed as

\begin{equation}\label{restmasscon}
(\rho u^{\mu})_{;\mu}=0 \,,
\end{equation}
where $u^{\mu}$ represents the four-velocity of the fluid, and $\rho$ is the rest mass density. The second equation is the radial component of the conservation of the energy-momentum tensor $T^{\mu \nu}$ aligned with the four-velocity

\begin{equation}\label{energycon}
u_{\mu} T^{\mu \nu}{}_{;\nu}=0\,.
\end{equation}
The third equation involves projecting the conservation of the energy-momentum tensor

\begin{equation}\label{NSE}
h_{\mu \sigma}(T^{\sigma \nu})_{;\nu}=0 \,,
\end{equation}
where $h^{\mu \nu} = u^{\mu} u^{\nu} + g^{\mu \nu}$ is the projection tensor. Considering the thin disk model's assumptions and applying them into the fundamental equations \eqref{restmasscon}-\eqref{NSE}, as well as incorporating the principles of radiative energy transport and vertical pressure gradients, we derive a set of nonlinear algebraic equations that describe the dynamics of thin accretion disks \citep{1973blho.conf..343N} as follows. 

According to the assumptions, heat flow is assumed to be in the vertical direction. Therefore, the time-averaged flux of radiant energy $F$, representing the energy per unit proper area and proper time, is emitted from the upper and lower surfaces and is directly related to this vertical heat flow \cite{1973blho.conf..343N,1974ApJ...191..499P}. 
By utilizing mentioned assumptions in the fundamental equations \eqref{restmasscon},\eqref{energycon}, and \eqref{NSE} we obtain
\begin{align}\label{ene}
\frac{(\Omega {L}-E)^2}{\Omega_{,{r}}}\frac{{F \sqrt{-|g|}}}{{\dot{M}}}= \int_{{r}_0}^{r} \frac{(\Omega {L}-{E})}{4\pi}{L}_{,{r}} d{r}, 
\end{align}
where $E=-u_t$ and $L=u_\phi$ are the energy and angular momentum per unit mass of geodesic circular motion in the equatorial plane, and $\Omega=\frac{u^{\phi}}{u^t}$ is the corresponding angular velocity. The radial velocity of the fluid is derived from the conservation of particle number
\begin{align}
u^r=-\frac{\dot{M}}{2\pi r \Sigma}. \label{massrate}
\end{align}
The next equation is the  surface density $\Sigma$ obtained by vertical integration of the density $\rho$
\begin{align}\label{sigma2}
\Sigma=\int^{+H}_{-H}\rho {\rm d}z =  2\rho H,
\end{align}
where $H$ is disk height or half of the thickness of the disk. The $\alpha$-prescription simplifies the complex physical processes by assuming that the viscous stress $W$ (vertically integrated) is proportional to the total pressure

\begin{equation}\label{w}
{W}= 2{\alpha}P {H},
\end{equation}
where $\alpha$ is a dimensionless parameter that represents the efficiency of angular momentum transport. The energy flux via viscosity reads as 
\begin{align}\label{navi}
F = -{\sigma_{\hat{\phi}\hat{r}}} {W}.
\end{align}
where $\sigma_{\hat{\phi}\hat{r}}$ is the off-diagonal part of the shear tensor, given by
\begin{equation}\label{viscosity}
\sigma_{\hat{\phi}\hat{r}} = \left(\frac{1}{2}(u_{\alpha ; \mu} h_{\beta}^{\mu} + u_{\beta ; \mu} h^{\mu}_{\alpha}) - \frac{1}{3} \theta h_{\alpha\beta}\right)\hat{e}_{r}\hat{e}_{\phi},
\end{equation}
The pressure equation in the vertical direction is given by \cite{1974ApJ...191..499P,1998bhad.conf.....K}
\begin{align}
    \frac{P}{\rho} = \frac{(HL)^2}{2r^4}
\end{align}
where the total pressure $P$ is the sum of the radiation pressure and the gas pressure, where

\begin{align}
    P_{\rm gas}& = \frac{\rho k_BT}{m_p}, \label{gasp}\\
    P_{\rm rad}& =\frac{a}{3}T^4.\label{radp}\
\end{align}
where ${m_p}$ is the rest mass of the proton, ${k_B}$ is Boltzmann's constant, $a$ is the radiation density constant, and $T$ is the temperature. Additionally, at each radius, since the emission is like black-body radiation, the energy transportation is given by
\begin{equation}\label{OD}
8\sigma {T}^4=3 {\Sigma} {F}{\kappa},
\end{equation}
where $\sigma$ is Stefan-Boltzmann's constant and $\kappa$ is Rosseland-mean opacity 
\begin{align}
    \kappa = \kappa_{\rm ff} + \kappa_{es},
\end{align}
where $\kappa_{\rm ff}$ is free-free absorption opacity and $\kappa_{es}$ is electron scattering opacity. Initially, $\kappa_{es}$ is equal to $\kappa_{es} = 0.2\times( 1 + X )$. Since the disk is dominated by the light element hydrogen, we approximate $X$ to be one. Therefore, we have 
\begin{align}
    \kappa_{es} =0.40 ~{\rm cm^2 g^{-1}},
\end{align}
and considering the mass absorption coefficient of pure Hydrogen plasma
\begin{align}
\kappa_{\rm ff} = 6.4\times10^{22} \left(\frac{\rho}{{\rm g \hspace{0.1cm}cm^{-3}}}\right)\left(\frac{T}{K}\right)^{-\frac{7}{2}}~ {\rm cm^2 g^{-1}}.
\end{align}
In practice, once the background spacetime is fixed (i.e., a choice of $\lambda$ in equation \eqref{eq:fTBImetric}, we proceed as follows. First, we determine the equatorial circular geodesics and compute the orbital quantities $\Omega(r)$, $E(r)$ and $L(r)$ from the metric. Second, imposing the standard zero-torque condition at the inner edge $r_0=r_{\rm ISCO}$, the Page-Thorne relation \eqref{ene} yields the radiative flux profile $F(r)$. Finally, for fixed disk parameters $(M,\alpha,\dot M)$, the remaining relations \eqref{massrate}-\eqref{OD} form a closed local algebraic system at each radius, which determines $\Sigma(r)$, $u^{r}(r)$, $H(r)$ (or $h(r)=H/r$), $P(r)$, $\rho(r)$, $T(r)$ and the viscous
stress $W(r)$, selecting the physical branch by $T>0$ and $P>0$.

% In summary, we have eight equations \eqref{sigma2}-\eqref{OD}, and eight unknown functions $F$, $\Sigma$, $W$, $u_r$, $h$, $P$, $\rho$ and $T$ all dependent on the radial coordinate. This system of equations is algebraic and admits a unique solution when the physical conditions $T > 0$, $P > 0$ are imposed. The solution depends on three free parameters $M$, $\alpha$, $\dot{M}$, in addition to the parameter of the underlying spacetime. 

In addition, the thin disk can be divided into different regions, allowing us to derive three distinct local solutions. These solutions can be obtained based on the dominance of gas pressure over radiation pressure, as well as the electron scattering opacity over the free-free absorption opacity of the disk. The three relevant local solutions are 
 
\begin{itemize}

\item  Inner region: $P \simeq P_{\rm rad}$ and $k\simeq k_{es}$.

\item Middle region: $P \simeq P_{\rm gas}$ and $k\simeq k_{es}$.

\item Outer region: $P \simeq P_{\rm gas}$ and $k\simeq k_{ff}$.
\end{itemize}
%%%%%%%%%%%%%%%%%%
 The inner region influences the qualitative characteristics of the global solution, which can be approximated through a piecewise construction of three local solutions, each patched together according to its range of validity. In the following section, we analyze the solutions of this model and explore the different regions within the TBI background.
%%%%%%%%%%%%%%%%%%%%%%%%%%%%%%%%%%%%%%%%%%%%%%%%%%%%%%%%%%%%%%%%%%%%%%%%%%%%%%%%%%%%%%%%%%%%%%

%%%%%%%%%%%%%%%%%%%%%%%%%%%%%%%%%%%%%%%%%%%%%%%%%%%%%%%%%%%%%%%%%%%%%%%%%%%%%%%%

\section{Results and discussion}\label{sec:Result}

In this section, we adopt the parameters
\begin{align}\label{param}
M&={M_{\odot}} \simeq 1.99 \times 10^{33} {\rm g},\nonumber\\
\alpha&=0.01,\nonumber\\
\dot{M}&=1.33\times10^{16} {\rm g \,s^{-1} } = 9.5 \times 10^{-2} L_{\rm Edd}/c^2.
\end{align}
With this choice, $\lambda$ can be compared directly to the Solar-mass weak-field bounds in equation \eqref{eq:lambdappn}, while results for other source masses follow from $\lambda(M)=(M/M_\odot)\lambda_\odot$.
We shall now study how the boundaries of different disk regions change with $\lambda$, and investigate the disk properties in those regions.    

%Aside from $h=\frac{H}{r}$, which is a dimensionless height scale, the results are represented in cgs units.
%% \begin{align}
%%m&=\frac{M}{M_{\odot}}\,,\\ \dot{m}&=\textcolor{blue}{\dot{M}c^2} \,,
%%\end{align}

\subsection{Different regions}

Based on the ratios of $P_{\rm gas}/P_{\rm rad}$ and $\kappa_{\rm ff}/\kappa_{\rm es}$, discussed in the previous section, we can distinguish between different regions in the disk. Figure \ref{fig:startingp-comparison3} examines how the radius at which $P_{\rm gas}/P_{\rm rad}=1$, which indicates the end of the inner region, changes with variations in $\lambda$, the viscosity coefficient $\alpha$ and the mass of the central object $M$. The results indicate that $\lambda$ has a relatively minor effect on this radius, while a change of one order of magnitude in $\alpha$ leads to a significant increase in the radius, particularly when the accretion rate is at the critical value. When the mass of the central object increases compared to the solar mass, the radius where $P_{\rm gas}/P_{\rm rad}=1$ is also larger. However, this deviation decreases as the accretion rate decreases. The effect of increasing $\lambda$ on the size of the inner region is opposite to that of increasing $\alpha$ and $M$. 
 
\begin{figure}
    \centering
    \includegraphics[width=1\hsize]{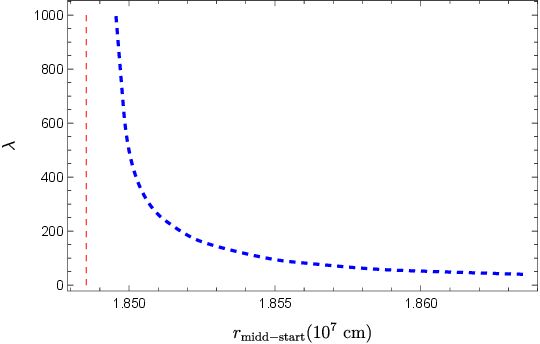}
    \includegraphics[width=1\hsize]{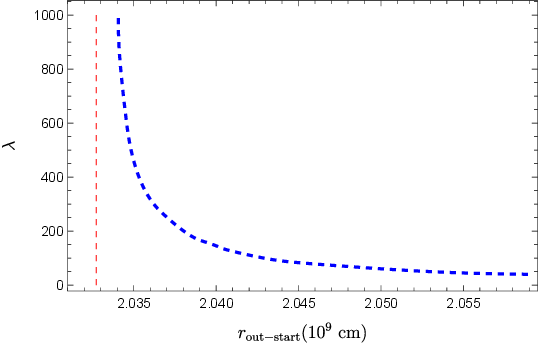}
    \caption{Starting points of the middle region (top plot) and outer region (bottom plot) as functions of $\lambda$. The dashed red line indicates the corresponding radii for the Schwarzschild solution. }  
    \label{fig:startingp-comparison}
\end{figure}

Figure \ref{fig:startingp-comparison} shows the starting point of the middle region and outer region as a function of $\lambda$ for the chosen mass accretion rate. The red dashed line serves as a baseline, highlighting the starting points in the Schwarzschild metric. The deviation of the blue curves from this line shows the degree to which $\lambda$ modifies the structure of the disk. In addition, the plots demonstrate that smaller values of $\lambda$ result in more extended radii for both the middle and outer regions. However, we should emphasize that the size of different regions is significantly influenced by the accretion rate.

\begin{figure}
    \includegraphics[width=\hsize]{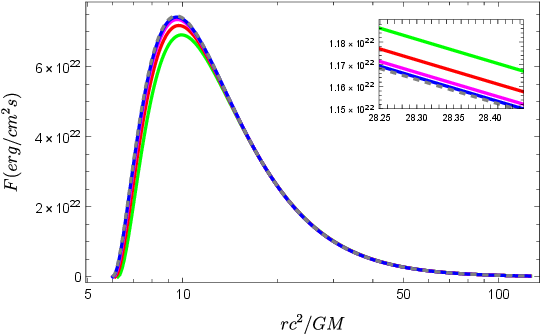}
    \includegraphics[width=\hsize]{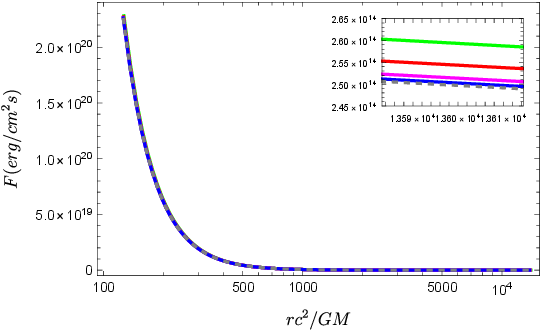}
    \includegraphics[width=\hsize]{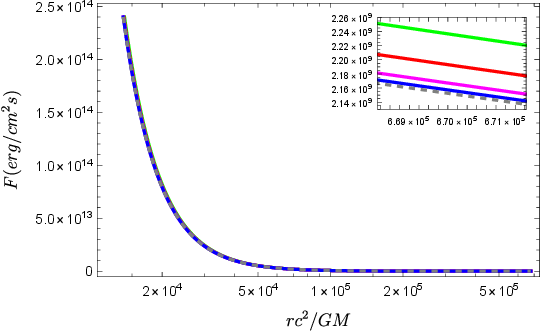}
    \caption{\label{fig:F-comparison3} Profile of flux, $F$ in three regions: inner (top), middle (middle), and outer (bottom), for the Schwarzschild (dashed line) and different values of $\lambda$ in TBI gravity: $\lambda = 40$ (green), $\lambda = 83$ (red), $\lambda = 231$ (pink), and $\lambda = 700$ (blue).}
\end{figure}

\subsection{Disk properties}

\begin{figure}
    \includegraphics[width=\hsize]{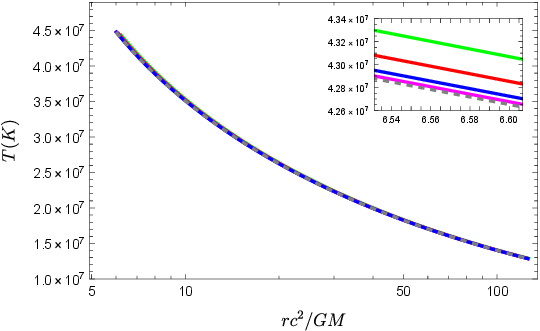}

    \includegraphics[width=0.97\hsize]{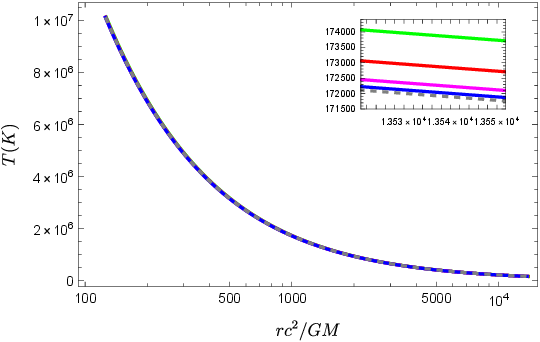}

    \includegraphics[width=0.97\hsize]{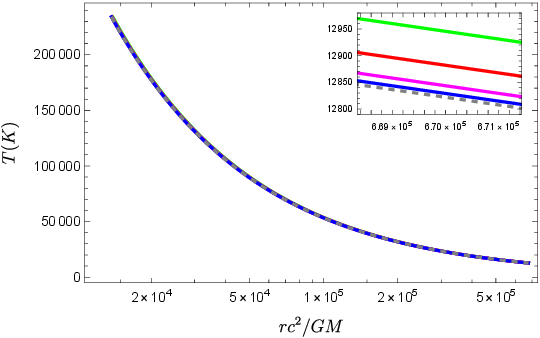}
    \caption{Temperature profile , $T$ in three regions: inner (top), middle (middle), and outer (bottom), for the Schwarzschild (dashed line) and different values of $\lambda$ in TBI gravity: $\lambda = 40$ (green), $\lambda = 83$ (red), $\lambda = 231$ (pink), and $\lambda = 700$ (blue).}
    \label{fig:T-comparison3}
\end{figure}
\begin{figure}
    \includegraphics[width=\hsize]{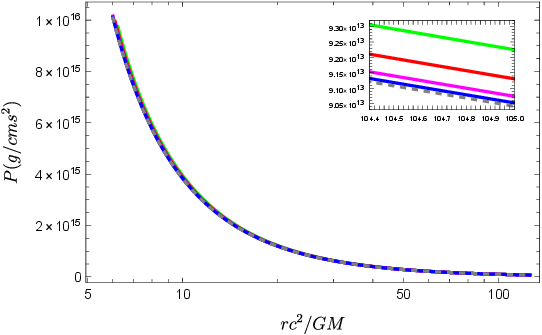}
    \includegraphics[width=\hsize]{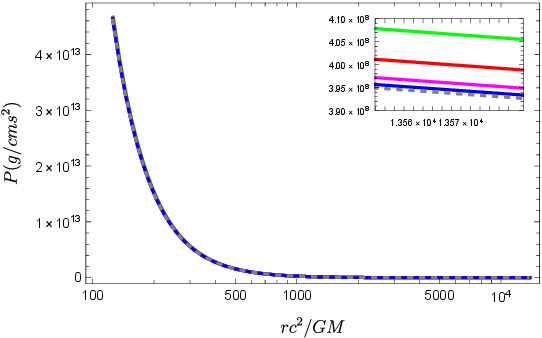}
    \includegraphics[width=\hsize]{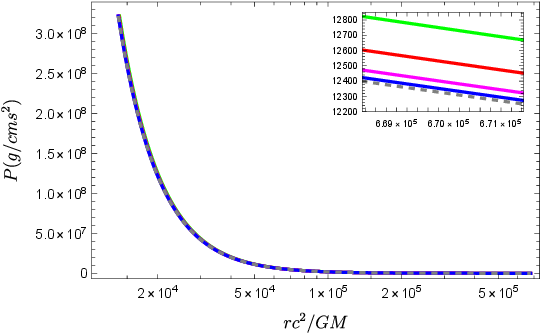}
    \caption{\label{fig:P-comparison3}Profile of pressure, $P$,
for different values of $\lambda$ and Schwarzschild in three regions: inner (top), middle (middle), and outer (bottom), for the Schwarzschild (dashed line) and different values of $\lambda$ in TBI gravity: $\lambda = 40$ (green), $\lambda = 83$ (red), $\lambda = 231$ (pink), and $\lambda = 700$ (blue).}
    \label{fig:P-comparison}
\end{figure}
\begin{figure}
    \includegraphics[width=\hsize]{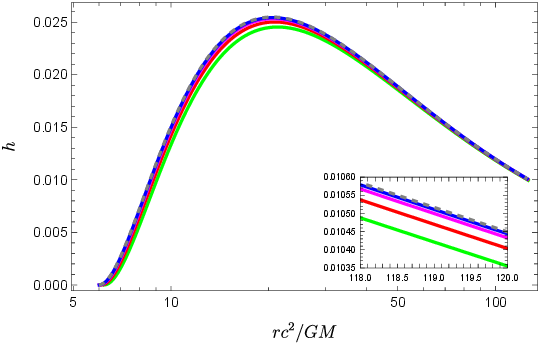}
    \includegraphics[width=0.96\hsize]{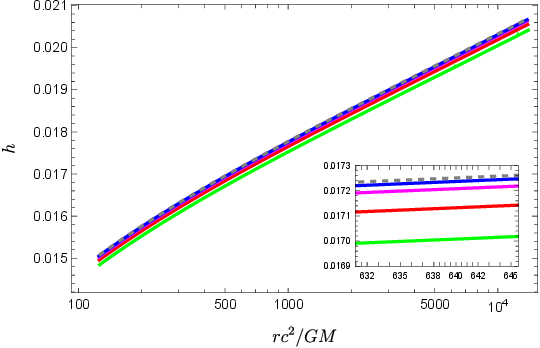}

    \includegraphics[width=0.96\hsize]{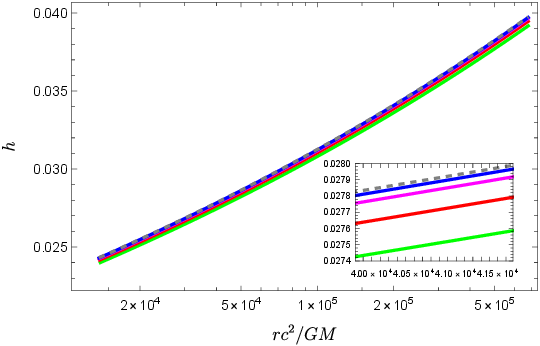}
    \caption{Profile of the disk height-to-radius ratio, $h$, for different values of 
 $\lambda$ and Schwarzschild in three regions: inner (top), middle (middle), and outer (bottom), for the Schwarzschild (dashed line) and different values of $\lambda$ in TBI gravity: $\lambda = 40$ (green), $\lambda = 83$ (red), $\lambda = 231$ (pink), and $\lambda = 700$ (blue).}
    \label{fig:h-comparison3}
\end{figure}
\begin{figure}

    \includegraphics[width=\hsize]{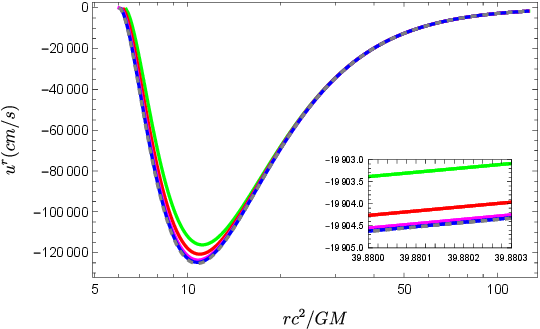}

    \includegraphics[width=0.95\hsize]{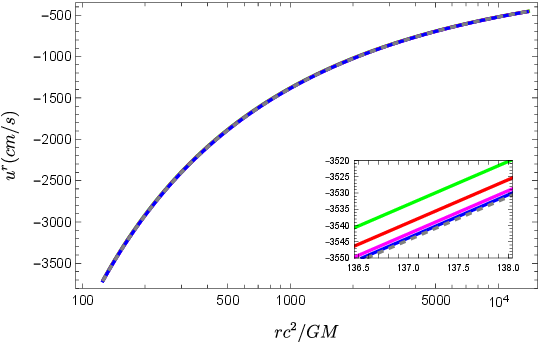}
    \includegraphics[width=0.95\hsize]{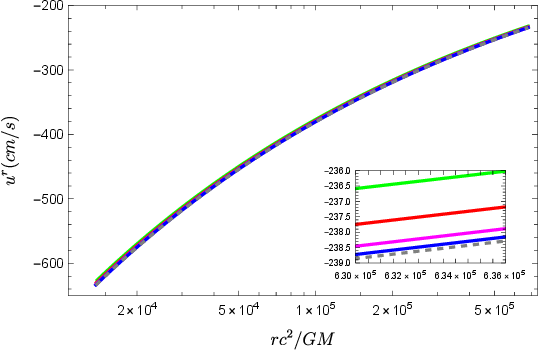}
    \caption{Profile of radial drift velocity, $u^r$, for different values of $\lambda$ and Schwarzschild in three regions: inner (top), middle (middle), and outer (bottom), for the Schwarzschild (dashed line) and different values of $\lambda$ in TBI gravity: $\lambda = 40$ (green), $\lambda = 83$ (red), $\lambda = 231$ (pink), and $\lambda = 700$ (blue).}
    \label{fig:ur-comparison3}
\end{figure}
\begin{figure}
%\begin{tabular}{ccc}
   % \includegraphics[width=0.7\hsize]{fig/W-in-final1.eps}&
    \includegraphics[width=\hsize]{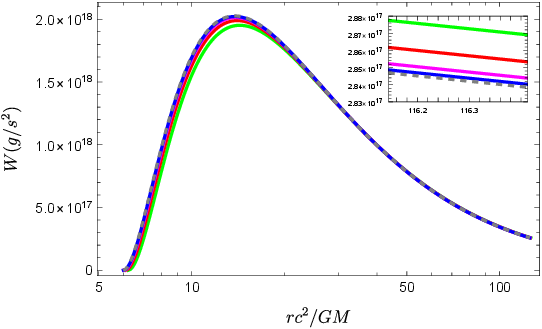}

    \includegraphics[width=\hsize]{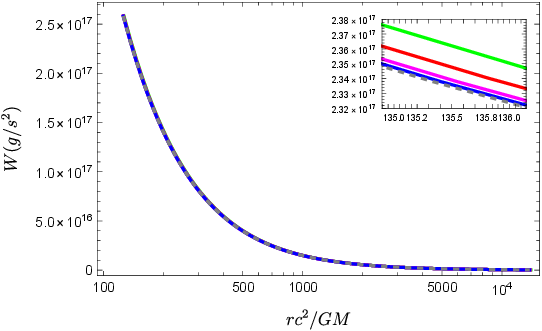}

    \includegraphics[width=\hsize]{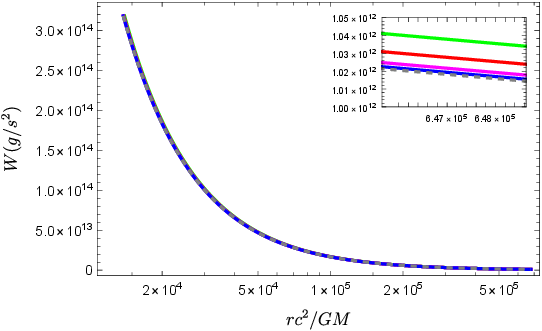}
   % \end{tabular}
    \caption{ Profile of viscous tensor, $W$, for different values of $\lambda$ and Schwarzschild in three regions: inner (top), middle (middle), and outer (bottom), for the Schwarzschild (dashed line) and different values of $\lambda$ in TBI gravity: $\lambda = 40$ (green), $\lambda = 83$ (red), $\lambda = 231$ (pink), and $\lambda = 700$ (blue).}
    \label{fig:w-comparison3}
\end{figure}

 We present the behaviour of different disk properties across three regions in Figures \ref{fig:F-comparison3}-\ref{fig:w-comparison3}. Figure \ref{fig:F-comparison3} presents the flux profile for various values of $\lambda$ comparing them with the Schwarzschild case and divided into three regions of the disk: inner, middle, and outer. We see that the flux in the inner region shows a prominent peak, indicating that this region has a significant flux contribution. The height and location of the peak vary slightly with $\lambda$; with smaller $\lambda$ values (e.g. green for $\lambda=40$) generally produce a lower maximum flux at a slightly larger radius. The middle region shows a declining flux with an increasing radial distance. As $\lambda$ decreases, the flux at a given radius increases, although the effect becomes much weaker at larger radii. In the outer region, the influence of $\lambda$ on the flux diminishes even further, making the flux relatively insensitive to changes in $\lambda$. 
 
 % Compared to the Schwarzschild case (the dashed gray line), $\lambda$ alters the flux, particularly by suppressing it closer to the central object, while enhancing at larger distances. 

To quantify this redistribution relative to the Schwarzschild case, we introduce the flux ratio
\begin{equation*}
\mathcal{R}_F(r;\lambda)\equiv \frac{F(r;\lambda)}{F_{\rm Schw}(r)} .
\end{equation*}
Denoting by $r_{\rm eq}(\lambda)$ the crossing radius where $\mathcal{R}_F(r_{\rm eq};\lambda)=1$
(i.e., the intersection of the finite-$\lambda$ and Schwarzschild curves in Figure \ref{fig:F-comparison3}),
the regions $\mathcal{R}_F<1$ and $\mathcal{R}_F>1$ provide a precise meaning of flux suppression
and enhancement, respectively.

Figure \ref{fig:T-comparison3} shows the (vertically-averaged) temperature profile in these three regions. In the thin disk model, the temperature and flux are related through Equation \eqref{OD}, which means that regions with higher temperatures will generally have higher flux values. Figure \ref{fig:T-comparison3} shows that as $\lambda$ decreases, the temperature in the disk is generally higher for a given radius. This higher temperature should result in a higher flux, consistent with the flux profile. Figure \ref{fig:P-comparison} shows the pressure. In the inner region, where the radiation pressure \eqref{radp} dominates, higher temperatures lead to significantly higher pressures, especially for smaller $\lambda$. As we move outward, the temperature decreases and thus the gas pressure takes over. In these regions, pressure $P$ decreases more gradually, as it scales linearly with temperature (Equation \ref{gasp}).

Figure \ref{fig:h-comparison3} represents $h=H/r$, where $H$ is the half-height of the disk.  In the inner region, $H$ increases as the radial distance increases. The disk height is smaller for smaller $\lambda$ (e.g., green for $\lambda=40$) and larger for larger This trend is opposite to what we saw in pressure, temperature, and flux, where smaller 
$\lambda$ led to higher values. Similar trends are observed in the middle and outer regions, $H$ continues to increase with radius, and the height of the disk is larger for larger values of $\lambda$. In conclusion, the smaller $\lambda$ values might lead to more compact, hotter, and denser disks, which could result in a thinner disk (smaller $H$). This could be because higher pressure and temperature in a more compact disk create a balance where the disk height is reduced.
On the other hand, the larger $\lambda$ values correspond to a more extended disk with a larger height, even though the temperature, pressure, and flux are lower. 

Figure \ref{fig:ur-comparison3} shows the radial drift velocity profile, $u^r$. In an accretion disk, matter typically drifts inwards due to the loss of angular momentum, so $u^r$ is usually negative.  The radial drift velocity is most negative in the inner region, where matter is rapidly falling inward. There is a clear minimum in the plot that indicates the radius where the inward drift is most significant. For smaller $\lambda$, this minimum is shallower, suggesting a slower inward drift compared to larger $\lambda$ values. As we move outward, the radial drift velocity becomes less negative, indicating a slower inward movement of matter in the cooler, less energetic outer parts of the disk. The effect of $\lambda$ diminishes in these regions, but the general trend remains that a smaller $\lambda$ leads to a slightly slower inward drift. The regions where the radial drift velocity is most negative (inner regions) correspond to the regions of highest temperature, flux, and pressure. This makes sense because in these regions, matter is rapidly losing angular momentum, falling inward, and heating up, leading to higher radiation (flux) and pressure. Figure \ref{fig:w-comparison3} presents the viscous stress tensor, $W$, related to the shear stress within the accretion disk and provides further insight into the energy dissipation in the disk. Considering the relation between $F$ and $W$ by equation \eqref{navi}, since the shear stress in the thin disk model varies slowly in the outer regions, the shape of the viscous stress $W$ closely mirrors that of the radiated flux $F$. However, in the inner region, $\sigma_{\hat{r}\hat{\phi}}$ varies steeply due to strong relativistic effects, leading to a wider peak for the viscous stress, $W$ compared to that of the flux, $F$.

\begin{figure*}
    \centering
    \includegraphics[width=0.5\hsize]{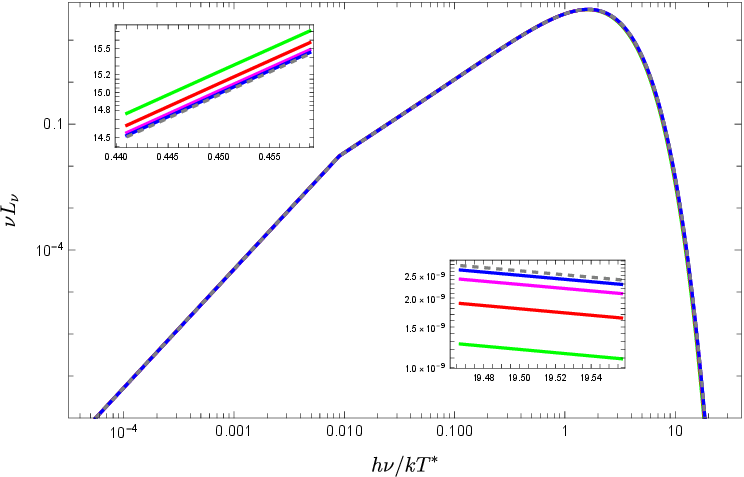}
    \caption{
Spectral luminosity $\nu \mathcal{L}_{\nu}$ in terms of the ratio of the photon energy $h\nu$ to the thermal energy $KT_{*}$ for the Schwarzschild case (dashed line) and for different values of $\lambda$ in TBI gravity: $\lambda = 40$ (green), $\lambda = 83$ (red), $\lambda = 231$ (pink), and $\lambda = 700$ (blue).}
    \label{fig:lum-label}
\end{figure*}

\begin{figure*}
    \centering
       \includegraphics[width=0.5\hsize]{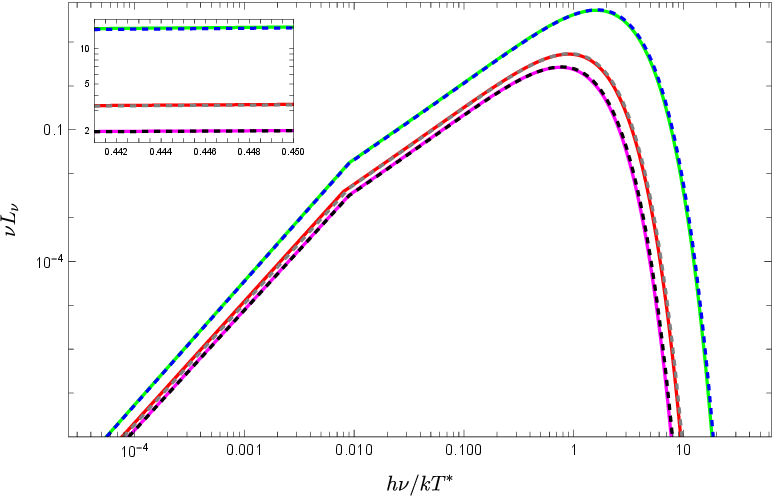}
    
     \caption{Comparison of different cases for $\lambda=40$ and Schwarzschild scenarios at varying mass accretion rates ($\dot{M}$) relative to the Eddington rate ($\dot{M}_{\text{Edd}}$). The solid green, red, and pink lines correspond to $\lambda =40$ with $\dot{M} = \dot{M}_{\text{Edd}}$, $0.1 \dot{M}_{\text{Edd}}$, and $0.05 \dot{M}_{\text{Edd}}$, respectively. The dashed blue, gray, and black lines represent the Schwarzschild case at the same $\dot{M}$ values.}
    \label{fig:lum-label2}
\end{figure*}

The differential luminosity,  $\frac{d \mathcal{L}_{\infty}}{d \ln r}$  that describes energy per unit time reaching the observer at infinity, can be approximated by flux through \cite{1973blho.conf..343N,1974ApJ...191..499P}
\begin{align}
    \frac{d \mathcal{L}_{\infty}}{d \ln r} = 4 \pi r \sqrt{g} E F.
\end{align}
Finally, assuming blackbody radiation, we can compute the spectral luminosity $\mathcal{L}_{\nu,\infty}$ in this spacetime, which is most directly related to observations of accretion disks e.g., \cite{2014SSRv..183..295M}. Spectral luminosity represents the amount of energy radiated per unit time per unit frequency interval \cite{1973blho.conf..343N}

\begin{align} \label{eq:SL}
    \nu \mathcal{L}_{\nu,\infty} = \frac{60}{\pi^3} \int_{r_{\rm ISCO}}^{\infty} \frac{\sqrt{g} E}{M^2} \frac{(u^t y)^4}{\exp\left[u^t y / F^{1/4}\right] - 1} dr,
\end{align}
is dimensionless quantity and $y=h_p\nu/k T_{*}$, $h_p$ is the Planck constant, $k$ is the Boltzmann constant, $T_{*} \equiv (\dot{M} c^2)/(4 \pi M^2 G^2 \sigma)$ is the characteristic disk temperature, and $\sigma$ is the Stefan-Boltzmann’s constant. 

\begin{figure*}
      \includegraphics[width=0.45\hsize]{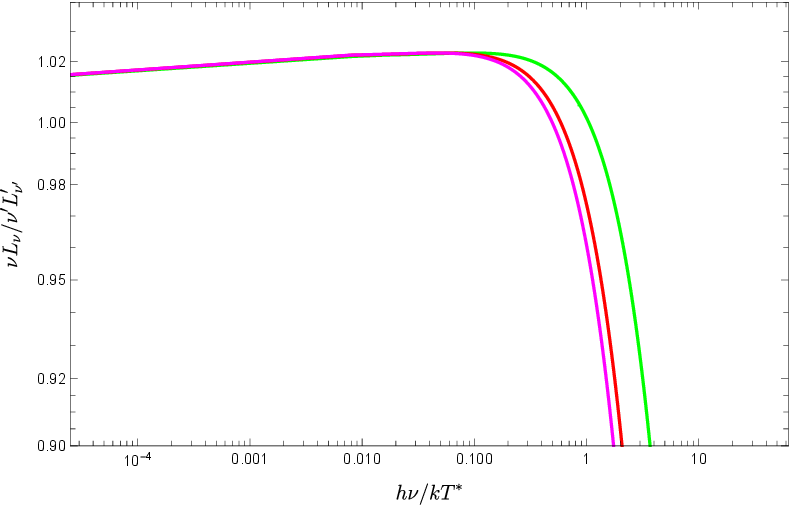}
     \includegraphics[width=0.45\hsize]{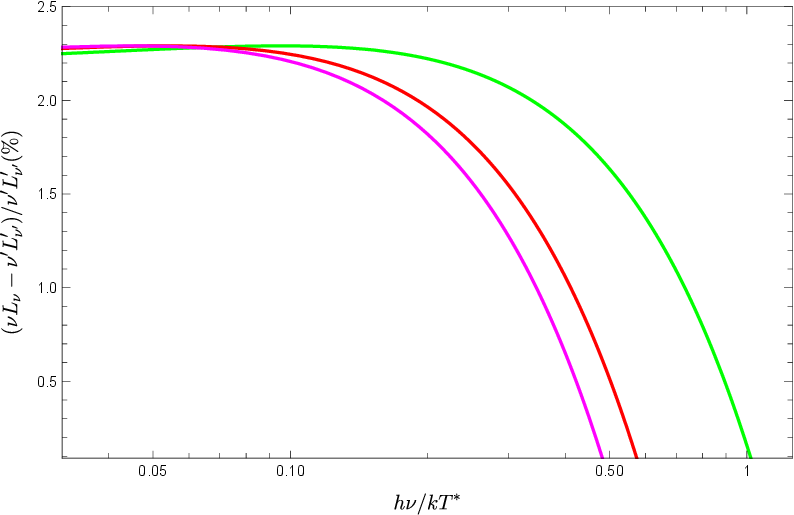}
          \caption{\label{ratioplot} Ratio of spectral luminosities for $\lambda = 40$ compared to the Schwarzschild case at different mass accretion rates ($\dot{M}$). The solid green line represents $\dot{M} = M_{\text{Edd}}$, the red line corresponds to $\dot{M} = 0.1 M_{\text{Edd}}$, and the pink line represents $\dot{M} = 0.05 M_{\text{Edd}}$. The lower plot displays the percentage differences.}
          %(LogLog shape)$(\lambda(40)-Schw)/Schw$\,*100\% (LogLinear shape for x-axies
  
\end{figure*}
This spectral luminosity formula \eqref{eq:SL} accounts for the contributions of all radii starting from ISCO outward to infinity, integrating the effects of gravitational redshift, energy, and relativistic beaming on the radiation emitted. The term $(u^t y)$ captures the relativistic effects of time dilation and the disk's velocity structure, particularly the influence of these factors on the emitted radiation. Figure \ref{fig:lum-label} shows how the spectral luminosity varies with the dimensionless frequency parameter for different values of $\lambda$ compared to Schwarzschild. Integration over $r$ from $r_{\rm ISCO}$ to infinity smooths out many of the variations that might exist at specific radii and reflects the cumulative effect of the entire disk. As a result, even though the conditions in the disk vary with $\lambda$, the overall spectral luminosity integrates these effects in a way that makes these deviations less visible. %In contrast, to the flux plot, where slight deviations were more apparent, the spectral luminosity integrates the effects over a broader range, making these deviations less visible. 
The curves closely overlap across the entire range of $h_p\nu/kT$ suggesting that the overall distribution of energy across frequencies remains consistent, even if the total energy output (flux) varies slightly with $\lambda$.

Figure \ref{fig:lum-label2} contrasts the effect of changing the accretion rate for a chosen value of $\lambda$. It shows that the spectral luminosity intensity increases monotonically with the mass accretion rate. Note that here we use the normalization $\dot M_{\rm Edd}\equiv L_{\rm Edd}/c^{2}$; for a radiatively efficient thin disk
$L\simeq \eta\,\dot M c^{2}$ with $\eta=1-E_{\rm ISCO}$, so that $\dot M=\dot M_{\rm Edd}$ corresponds to $L\simeq \eta L_{\rm Edd}$ (e.g. $\eta\simeq 0.057$ for Schwarzschild), well within the thin-disk regime.
As the accretion rate decreases the peak luminosity drops significantly, and the overall curve shifts downward. This suggests that as the accretion rate decreases, the radiative energy output shifts towards lower frequencies, resulting in a less energetic spectrum. %The overall trend and flipping behaviour remain unchanged across different accretion rates.
Figure \ref{ratioplot} illustrates the ratio of spectral luminosities for $\lambda=40$ compared to the Schwarzschild case for different mass accretion rates. The upper plot shows the ratio of the spectral luminosities, while the lower plot presents the percentage differences between the same models. In the ratio plot, for photon energies $h_p\nu/kT_* < 1$, the curves for all three mass accretion rates remain relatively close to 1, indicating that the spectral luminosities for the $\lambda = 40$ model are very similar to those for the Schwarzschild case. However, as seen in the percentage difference plot, the differences remain small, on the order of a few percent, even when the curves in the ratio plot appear nearly indistinguishable. Given the current sensitivity of X-ray observations, particularly near the spectral peak (e.g., \cite{2014SSRv..183..295M}), it may be possible to detect such differences under favorable observational conditions. This suggests that even slight variations in the models could potentially be distinguished through careful data fitting, especially if the observed spectra cover a wide range of frequencies. Although the differences are subtle, they are critical for understanding how modifications to gravity affect thin disk accretion models. They also provide a theoretical foundation for exploring how the parameter $\lambda$ influences observable features, extending the study of accretion processes beyond the Schwarzschild framework. 

\begin{figure}
    \centering
        \includegraphics[width=0.95\hsize]{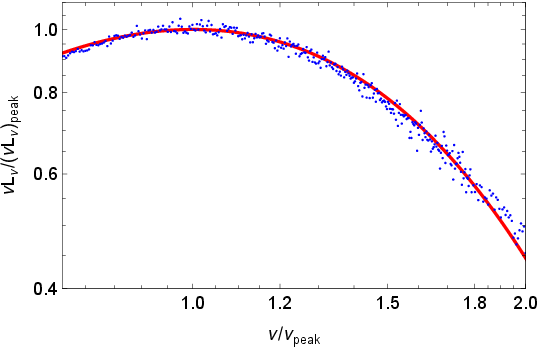}
    \includegraphics[width=0.95\hsize]{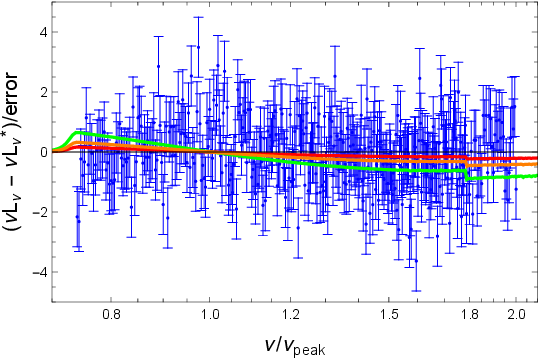}
     \includegraphics[width=0.95\hsize]{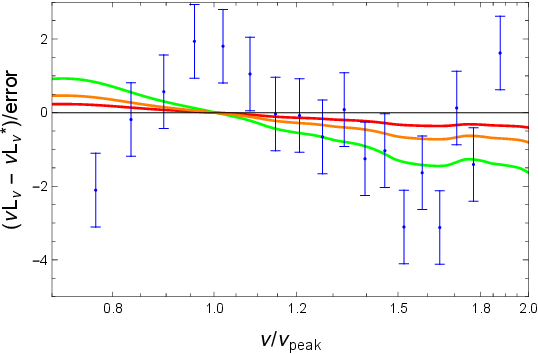}
         \caption{Top panel: Theoretical model fit for $\lambda=400$ (red line) compared to observational data (blue points). \\ 
         Middle panel: Unbinned residuals showing the scatter in individual data points, and confirming that there are no significant systematic trends. Curves show changes relative to general relativity, with $\lambda=100$ (green), $\lambda=200$ (orange), and $\lambda =400$. \\          
         Bottom panel: Binned residuals with errorbars representing statsitical error on the mean within each bin. 
         \\
         The mass accretion rate is given by $\dot{M}=\dot{M}_{\text{Edd}}$. }
    \label{fig:data}
\end{figure}

The top panel of Figure \ref{fig:data} compares the prediction of the theoretical model for $\lambda=400$ to the observational data for the low-spin X-ray black hole binary MAXI $J1820 + 070$ \cite{10.1093/mnras/stab945} which is consistent with Eq. (2.3) \ref{eq:fTBImetric}. The model successfully captures the overall shape and key features of the data within the range $\nu/\nu_{\rm peak}\leq 2$, particularly near the peak, where the agreement is strongest. This shows the model’s ability to describe the physical processes relevant to the low-frequency regime. However, we limit our analysis to $\nu/\nu_{\rm peak}\leq 2$ as the model does not account for additional physical mechanisms that dominate at higher frequencies. Addressing the high-frequency regime would require extending the model, for instance, with a power-law component, to account for the observed trends.

The middle panel of Figure \ref{fig:data} presents the unbinned residuals, offering a detailed view of the individual deviations between the model and the data. We notice that the models with different $\lambda \gtrsim 100$ are all broadly consistent with these residuals.   

Finally, the binned residuals (bottom panel of Figure \ref{fig:data}) provide a direct estimate of the statistical leverage of current data on the finite-$\lambda$ spectral distortions in the
band $\nu/\nu_{\rm peak}\le 2$. In this range, the expected deviations from the Schwarzschild
spectrum are ${\cal O}(1)$ in units of the observational uncertainty for $\lambda\sim 10^2$, and
drop below the sub-$\sigma$ level for $\lambda \gtrsim 400$. This motivates a dedicated continuum-fitting
analysis that marginalizes over the usual disk and system parameters to translate this sensitivity
into a robust constraint on $\lambda$ (or equivalently on $\tilde{\Lambda}$).

%The ratio decreases as $h\nu/KT_*$  increases, indicating that the spectral luminosity for smaller $\lambda$ differs more significantly from the Schwarzschild case at lower mass accretion rates and higher frequencies. At the Eddington accretion rate, the curves show a gradual decline with relatively small deviations. However, as the accretion rate decreases, the slope of the curves steepens significantly, indicating that the spectral luminosity for smaller $\lambda$  diverges more sharply from the Schwarzschild case, particularly at higher frequencies. 

%the third one is the previous error $\pm 0.5 + \text{binning operation error}$
%\caption{\textcolor{blue}{The top panel is unbinned one. The latter plots set the bin as 20 and $\lambda=400$ in red, $\lambda=200$ in orange, $\lambda=100$ in green. The second one is: the binning operation error ($\sigma/\sqrt{20}$) and  The fourth one is $\sqrt{\left(\left(\frac{\sigma}{\sqrt{20}}\right)^2 + 0.5^2\right)}$.}} \label{fig:enter-label}
%%%%%%%%%%%%%%%%%%%%
\section{Summary and conclusion}\label{sum}
Born-Infeld teleparallel gravity is one of the most successful f(T)-gravity theories so far, at least in terms of the existence of non-perturbative solutions. In this paper, we proposed a different route to test its viability and paved the way for further investigations by studying the standard thin-disk model in this modified theory of gravity.

In particular, in this study, we investigated the impact of the spacetime parameter $\lambda$ on the physical properties of thin accretion disks, focusing on quantities such as flux, pressure, temperature, etc. The findings reveal that $\lambda$ influences the structure and dynamics of the disk, particularly in the inner regions. Smaller values lead to hotter, more compact disks with increased flux, higher radiation pressure and a reduced magnitude of the inward drift velocity. These effects are especially pronounced near the ISCO. This suggests that $\lambda$ not only affects the local conditions (pressure, temperature, flux) but also plays a role in determining the overall structure and geometry of the disk. Overall, $\lambda$ emerges as a critical parameter in determining the physical characteristics of thin disks, with implications for both theoretical models and observational studies of accreting black holes and neutron stars. 

We also examined the spectral luminosity in this spacetime, comparing different $\lambda$ values and the Schwarzschild solution. The results show that while the curves are similar at higher accretion rates, differences become more pronounced at lower accretion rates and higher frequencies. This indicates that $\lambda$ has a greater impact in low luminosity systems, where deviations from the Schwarzschild model are more noticeable and can help distinguish between different $\lambda$ models and Schwarzschild. 

The inclusion of observational data further demonstrates the model's capability to align with real astrophysical systems. By comparing the theoretical predictions with data in the low-frequency regime, we highlight the potential to distinguish subtle differences between modified gravity theories and general relativity. 

Future research could explore how $\lambda$ affects more complex disk phenomena, such as interactions with magnetic fields, and how these effects could manifest in observable systems. These studies would provide valuable insights into the differences between modified gravity theories and general relativity. Additionally, further efforts to compare theoretical predictions with broader datasets could refine our understanding of how modified gravity aligns with observations, offering a more concrete basis for testing these scenarios.

%\begin{figure}
%    \centering
%      \includegraphics[width=\hsize]{fig/Unbinedone.eps}
 %        \caption{\textcolor{blue}{Fig1: lum $\lambda=400$ is red line and blue dots are data. Fig2: $\lambda=100$ is green line, $\lambda=200$ is orange line and $\lambda=400$ is red line. Fig3: Red is $\lambda=400$ and blue is data. Fig4: Red dot is $\lambda=400$, orange is $\lambda=200$ and green is $\lambda=100$}}
 %   \label{fig:enter-label}
%\end{figure}

%%%%%%%%%%%%%%%%%%%%%%%%%%%%%%%%%%%%%%%%%%%%%%%%%%%%%%%
 \begin{acknowledgments}
  The authors thank James Steiner for fruitful discussions. This research is supported in part by the University of Waterloo, Natural Sciences and Engineering Research Council of Canada, by the Government of Canada through the Department of Innovation, Science and Economic Development and by the Province of Ontario through the Ministry of Colleges and Universities at Perimeter Institute. 
 \end{acknowledgments}

\bibliographystyle{unsrt}
\bibliography{bibbornfinal}

\end{document}